# Influence of morphological inhomogeneity induced carrier diffusion on transient photocurrent pulse shape in organic thin films

S. Raj Mohan[*,1], Manoranjan P. Singh[2], M. P. Joshi[1], L. M. Kukreja[1]

[1]Laser Materials Processing Division, [2]Laser Plasma Division,

Raja Ramanna Centre for Advanced Technology, Madhya Pradesh, Indore, India, 452013.

## Abstract

The influence of film morphology induced carrier diffusion on the broadening of the time-of-flight transient photo-current pulse was investigated using Monte Carlo simulation in organic thin films. Assuming the Gaussian Disorder Model for the charge transport the simulation of the time-of-flight photo-current pulse shape was carried out for homogeneous and inhomogeneous films by varying the overall energetic disorder of the system. In the case of homogeneous system, the value of the tail broadening parameter ($W$) of the photocurrent pulse is found to decrease upon decreasing the energetic disorder. The observed behavior is explained by using the temporal evolution of carrier diffusion coefficient. In case of the inhomogeneous system, upon decreasing the overall energetic disorder of the system the value of $W$ initially attained a maximum before it started to decrease. This is attributed to the morphology dependent carrier diffusion in the latter case. This study elicits the importance of the influence of the film morphology induced carrier diffusion on the experimentally measured shape of the time-of-flight transient photo-current pulses, which is found to be generally ignored.

---

[*] Corresponding author. Tel.: +91 7312488304; fax: +91 7312488300
E-mail address: raj@rrcat.gov.in; rajmho@gmail.com (S. Raj Mohan).

**Introduction**

Extensive investigations on charge transport in disordered organic materials unambiguously establish the hopping charge transport in these materials resulting in low charge carrier mobility [1-5]. The common practice for experimental/theoretical investigation of charge transport in such materials is to study the dependence of charge carrier mobility on various parameters such as electric field, temperature, carrier concentration, film morphology etc [1-9]. Numerous experimental techniques have been adopted for measuring carrier mobility out of which the time-of-flight transient photoconductivity (TOF) experiment is quite common and is being actively pursued [2,4]. Main reasons for the wider acclaim of TOF experiment are (i) it is simple to perform (ii) the experimental results are independent of electrical contacts used and (iii) it can provide electron and hole mobility separately. In TOF experiment, thick organic layer is sandwiched between two metal electrodes, and thin sheet of charge carriers is generated (either by short laser pulse [1-4] or electron beam [10,11]) at one of the metal organic interface. The displacement current due to this thin sheet of charge carriers under a known applied potential is recorded as function of time. The shape of typical TOF current pulse consists of an initial spike, a plateau where the current/carrier drift velocity remains almost constant followed by a long tail which suggests the spreading of the sheet of charge carriers [1-4]. Charge carrier mobility can be calculated by knowing the thickness of the sample, the applied potential and the transit time of the sheet of carriers across the sample thickness [1-4]. Conventionally, the transit time is taken as the time at the point of intersection of tangents drawn on to the plateau and the falling tail part of the current pulse (referred as $t_0$) or the time at which the current has reduced to half of the value at the plateau region (referred as $t_{1/2}$) [1-4]. In some cases where a clear plateau is not observed, TOF current pulse on a double logarithmic scale is considered for extracting the transit time [1-4]. Even though the prime importance of the TOF experiment is to measure the drift mobility of electron and hole, but it also carry information related to trap states, density of states (DOS), carrier diffusion, recombination processes etc [12-16]. The parameters extracted out of study of TOF current pulse and the proposed models of charge transport are highly sensitive to the shape of the current pulse [12-16]. Any change in the experimental parameters/nature of the sample, that affects the behavior of charge transport, can significantly influence the shape of TOF current pulse [17-19]. Hence, a proper understanding of the various factors and mechanisms that influence the shape of TOF current pulse is important for extracting meaningful information about the sample. One of the experimentally measurable parameter which characterizes the shape of TOF current pulse is the tail broadening parameter $W$ which is defined as $W=(t_{1/2}- t_0)/t_{1/2}$ [17-19]. The parameter $W$ is related to the spreading or diffusion of sheet of charge carriers and hence depends on various material parameters such as energetic disorder, sample thickness, temperature etc. [17-19]. Therefore, a detailed investigation on the influence of various material parameters on the shape of TOF current pulse provides a deeper insight in to the mechanisms of charge transport.

Film morphology is known to have strong influence on the carrier transport. Several reports provide evidence of enhanced charge carrier mobility when structural orders (such as nano/micro scale ordering of polymer chains or molecular aggregates) are introduced in otherwise amorphous or disordered organic thin films [20-25]. In terms of energetic disorder, these morphologically tailored thin films are therefore not spatially homogeneous rather they are inhomogeneous [20,21]. The charge transport in such systems then occurs through a mixture of ordered and less ordered regions. In such systems, the energetic disorder seen by the carrier keeps fluctuating and hence influence of energetic disorder on charge transport becomes complex [26,27]. Recently we have made simulation studies on carrier diffusion in such inhomogeneous films and observed morphology dependent carrier diffusion which acts in addition to the thermal and non-thermal field assisted diffusion [28]. This additional carrier diffusion arises due to the spatial fluctuations in the energetic disorder and is attributed to the slow relaxation of the carriers generated in the less disordered regions present in the inhomogeneous system [28].

In this report, simulation studies are carried out to understand the influence of morphology dependent carrier diffusion on the shape of TOF current pulse. In most of earlier reports on simulation of photocurrent pulse shape, only homogenous nature of film was assumed [4,19,29-31]. The morphology of the sample in the present case was varied by randomly embedding ordered regions of less energetic disorder inside a host material of high energetic disorder (referred to as *inhomogeneous system* here after). Simulation is also performed without any embedded ordered regions (referred to as *homogeneous system* hereafter). The value of $W$ is obtained with various concentrations of ordered regions. It is known from earlier reports [22-24] that the overall energetic disorder seen by the carrier decreases with increase in the concentration of ordered regions. *W* is known to decrease with decrease of energetic disorder for a homogeneous system [18]. Hence, *W* is expected to decrease with increase of concentration of ordered regions. Contradictory to this expectation, we find that *W* initially increases and reaches a maximum value with increase of concentration of ordered regions beyond which it starts decreasing with further increase in the concentration of the ordered regions. The observed variation in *W* with increase in concentration of ordered regions is explained using morphology dependent carrier diffusion [28]. Thus, this study highlights the influence of morphology dependent carrier diffusion on experimentally observed results.

**Details of Simulation**

A 3D array with size 70x70x10000 along *x*, *y* and *z* direction is considered as the lattice. The size of the lattice is judged on the basis of our intention to change the lattice morphology and also by taking into account of the available computational resources. *Z* direction is taken as the direction of the applied field. The lattice constant $a = 6\text{Å}$ is used for the whole set of simulation [4, 22-24]. The site energies are assumed to be correlated and follow Gaussian distribution [4,23,24] with a known standard deviation ($\sigma$, the energetic disorder parameter). Simulation is performed on an energetically disordered lattice with the assumption that the hopping among the lattice sites is controlled by Miller-Abrahams equation [4,22-24] in which the jump rate ($\upsilon_{ij}$) of the charge carrier from the site *i* to site *j* is given by,

$$\upsilon_{ij} = \upsilon_o \exp\left[-2\gamma a \frac{\Delta R_{ij}}{a}\right] \begin{bmatrix} \exp\left[-\frac{\varepsilon'_j - \varepsilon'_i}{kT}\right] & , & \varepsilon'_j > \varepsilon'_i \\ 1 & , & \varepsilon'_i > \varepsilon'_j \end{bmatrix} \quad (1)$$

where $\Delta R_{ij} = |R_i - R_j|$ is the distance between sites *i* and *j*, $\varepsilon'_i$ and $\varepsilon'_j$ are the effective energies of the site *i* and *j* including the electrostatic energy, *a* is the intersite distance, *k* is the Boltzmann constant, *T* is the temperature in Kelvin and *2γa* is the wave function overlap parameter which controls the electronic exchange interaction between sites. Throughout the simulation the positional disorder is neglected with the value of overlap parameter, *2γa*=10 [4,22-24]. Simulation is based on single carrier approach which is valid for very low carrier concentration where the influence of space charge effects [32,33] can be neglected. We note here that TOF experiments are generally performed at very low carrier concentration, ($Q_{generated}<<0.05CV$) which supports the use of single carrier approach. The charge carrier is injected randomly on to the first plane of the lattice, which is then allowed to hop in the presence of applied electric field. In this report, simulations are carried out for E=6.4x10$^5$V/cm and T=300K. Every injected carrier is allowed to hop until it covers a sample length of *6μm*. Simulation is performed by averaging over ten thousand carriers with one lattice realization per carrier. Position of the carrier as a function of time is recorded from which the current pulse is calculated using *I(t)=eN(t)(<z(t)>/t)*, where *<z(t)>* and *N(t)* are the average position of carriers and the number of carrier contributing to the photocurrent at a time *t* respectively [31,34]. Transit time parameter $t_o$ and $t_{1/2}$ are extracted from the photocurrent pulse and used for calculating the value of *W* as explained above. Photocurrent is simulated for various values of energetic disorder and the value of *W* for each case is calculated. Knowing the position of the carrier as a function of time the temporal evolution of diffusion is also calculated using equation,

$$D_z = \frac{\left\langle \left(z - \langle z \rangle\right)^2 \right\rangle}{2t} \qquad (2)$$

Diffusion coefficient ($D_z$) presented in the manuscript is calculated along the applied field direction at a temperature T=300K. According to the recent report [35] the deep site energies around -2σ$^2$/kT are determinant for field dependent diffusion and hence very important. The report suggests that to have the required site energies (energies around ~-6σ for T=300K) in the simulation a lattice of much bigger size is essential. Our investigation on the need of bigger lattice size (details provided in supplementary information) suggests that lattice size chosen in this work (e.g. 70x70x10000) is sufficient to obtain accurate value of $D_z$ provided averaging is carried over large number of lattice realizations. The averaging over large number of lattice realizations will bring in the influence of deep site energies and hence accurate values of $D_z$ are obtained using smaller lattice sizes. Our investigations have shown that with the averaging carried over ten thousand carriers with one lattice realization per carrier used in the present case brings out the influence of required deep site energies [35].

We adopted a simplified model to investigate the influence of film morphology on the TOF current pulse. For inhomogeneous system the lattice morphology is varied by embedding cuboids of ordered regions inside a highly disordered host lattice [22-24]. The size and the location of cuboids of ordered region was random. Energetic disorder inside the ordered region is intentionally kept low compared to the host lattice. This assumption of high disordered lattice with embedded ordered regions is appropriate because in most practical devices the organic/polymer films employed are morphologically tailored. Therefore these organic films contains regions of low and high energetic disorder [20,22,23]. Energetic disorder for the host lattice is assigned as 75meV, (a typical value of energetic disorder seen in the homogeneous disordered organic materials), and for the ordered regions we chose it five times less compared to host lattice i.e. 15meV. Due to ordering of polymer chains or molecular aggregates a reduced value of σ is expected [23]. Earlier reports have even suggested a 10-fold reduction of energetic disorder inside the polycrystalline regions [36]. Mean energy of DOS for the host lattice and ordered regions are assumed to be same. Sizes of ordered regions are limited to a maximum of 25x25x40 sites along *x, y,* and *z* directions. Size of cuboids is chosen such that morphology can be varied upon changing the concentration of ordered regions. Moreover, random embedding of such cuboids of varying sizes in the host lattice can generate nano-scale morphology in the host lattice. Thus, this kind of lattice mimics the inhomogeneous case where the charge transport occurs through regions of high and low energetic disorder. Thus, the simulation brings in exclusively the influence of spatial fluctuations in the energetic disorder. As explained above, the photocurrent pulse is simulated by varying the concentration of ordered regions and the respective values of $W$ are calculated. Similarly, the temporal evolution of $D_z$ is calculated.

**Results and Discussions**

Figure 1 shows simulated TOF current pulses for various values of energetic disorder in homogeneous system. Current pulses are normalized to the value of current at the plateau region. All simulated current pulse exhibit an initial spike and a plateau region (with constant current) followed by a tail. The values of $t_0$ and $t_{1/2}$ are extracted from the current pulse as explained before. As expected for a homogeneous system, it is observed that charge carrier transit time decreases with decrease in the value of energetic disorder [4]. Consequently, mobility increases as evident from the inset of Figure 2. Using the values of $t_0$ and $t_{1/2}$ the value of $W$ is calculated for each case. As shown in the figure 2, $W$ decreases with decrease in the energetic disorder. Figure 3 shows the plots of $D_z$ as a function of time *(normalized to the dwell time of a lattice without disorder, τ)* for various values of energetic disorder. In all the cases, $D_z$ decreases initially and reaches a minimum value (valley point) before it starts increasing with time. The time at which valley point is observed decrease with the decrease in the energetic disorder, which provides another indication of the decrease in the overall energetic disorder seen by the carrier. This is related to the time taken by carriers to relax to the bottom of DOS [3,4,29,37]. After attaining a valley point, $D_z$ increase at much slower rate with decrease in energetic disorder. Also, after covering a given sample length, the maximum value of $D_z$ attained with respect to the valley point decrease with decrease in the energetic disorder. $D_z$ contains the contribution from both the thermal and the non-thermal field assisted diffusion. The non-thermal component is field and energetic disorder dependent, which arises due to the large difference in the jump rates of carriers occupied at the bottom and top of DOS [29,38]. Higher the value of energetic disorder higher is its contribution [29,38]. Therefore, after attaining minimum value the rate at which $D_z$ increase with time is expected to be high for large values of energetic disorder. In addition, the time at which steady state diffusion is attained (the constant value of $D_z$ with time) decreases with decrease in energetic disorder.

For inhomogeneous system, the simulated TOF current pulses as a function of concentration of ordered region is shown in Figure 4. Features of pulse shape are similar to homogeneous system i.e. initial spike, plateau region and followed by a tail. $W$ for each pulse is extracted and plotted in Figure 5. With increase in the concentration of ordered regions, it is expected that there would be decrease in the overall or effective energetic disorder of the system [22-24]. This is very much evident from the corresponding increase in the carrier mobility (see the inset of Figure 5). Therefore, one may expect a decrease in $W$ with increase in the concentration of ordered regions. Contrary to this expectation, we observe that $W$ increases with the increase in the concentration of ordered regions and attains a maximum at ~85% concentration of ordered region. Thereafter it decreases with further increase in the concentration of the ordered regions. To understand the behavior of $W$ in Figure 5, it is important to analyze the temporal evolution of diffusion of carriers in detail. Figure 6 shows the temporal evolution of $D_z$ for various concentrations of ordered regions. For comparison, $D_z$ for homogeneous system (0% and 100% concentration) is also shown in Figure 6. For all concentrations of ordered regions, it is observed that $D_z$ initially decreases and reaches a minimum (valley point) and then it begins to increase with time. The time at which the valley point is reached, decrease with increase in the concentration of the ordered regions. This once again suggests that as the concentration of ordered regions increases the overall energetic disorder of inhomogeneous system decrease. At low concentration of ordered region (for eg. 20%, 40%, 60%), after attaining the valley point the value of $D_z$ increases with time (the initial increasing regime) and then reaches an regime (the intermediate regime) where the rate of increase in $D_z$ is small. After the intermediate regime, the $D_z$ increases further with time at a rate lower than the initial increasing regime but higher compared to intermediate regime. Rate of increase of $D_z$ after the intermediate regime is approximately same as that of pure host lattice. These three different regimes of different slope are clearly seen in Figure 6 e.g. for 20% concentration. The intermediate regime gradually vanishes upon increasing the concentration of ordered region. Once the intermediate regime is vanished, at higher concentration (~≥80%), increase of $D_z$ after the initial increasing regime is approximately same as that of pure host lattice. In the initial increasing regime, the rate of increase of $D_z$ increases up to an optimum concentration of ordered region. The rate attains a maximum value at ~85% concentration of ordered regions beyond which it decreases with further increase in the concentration of ordered regions. Exactly the same trend is observed for the magnitude of $D_z$ measured with respect to the valley point. Compared to homogeneous system the temporal evolution of $D_z$ in the inhomogeneous system is quiet intriguing. Time evolution of $D_z$ for inhomogeneous system can be explained using morphology dependent carrier spreading mechanism [28] that acts on the carrier packet in addition to the thermal and non-thermal field assisted diffusion. The morphology dependent carrier spreading mechanism arises due to the slow relaxation of the carriers generated in the ordered regions of inhomogeneous system [28]. The behavior of $W$ for inhomogeneous system can be explained in the same way as homogeneous system. Higher the rate of increase in the $D_z$ in the initial increasing regime, higher is the spreading in the carrier packet and hence a larger $W$. The gradual vanishing of the intermediate plateau regime with increase in concentration of ordered region also results in the increase in value of $W$ which attains the maximum value at ~ 85% concentration of ordered regions. It should be noted that $W$ attains the maximum value approximately at the same concentration for which the rate of increase of $D_z$ and its magnitude with respect to the valley point is maximum. Further increase in concentration of ordered regions results in the decrease in $W$. Moreover, as the concentration of ordered region increases the whole lattice tend towards the homogeneous system with low disorder, which also supports the decreases of $W$. In concise, the observed behavior of $W$ upon increasing the concentration of ordered region can be explained using temporal evolution of $D_z$ that arises due to the morphology dependent carrier spreading.

**Conclusions**

In conclusion, this study establishes the influence of morphology on the shape of TOF current pulse and in particular highlights the influence of morphology dependent carrier diffusion on the experimental observations and extracted parameters. In organic photonic devices like organic solar cells, organic field effect transistors etc the active layers are often not truly amorphous. In such cases, the morphology dependent carrier diffusion can significantly influence exciton diffusion, recombination, charge transport and other processes. Thus, understanding the influence of morphology dependent carrier diffusion on the performance of the device is very important. This study not only establishes the need to consider the film morphology while modeling the TOF pulse shape but also highlights the relevance of morphology dependent carrier spreading mechanism in understanding the organic photonic devices better and in optimizing their performance.

**Appendix A. Supplementary material**

Supplementary data associated with this article can be found in the online version.

**Figure captions**

Figure 1. Simulated TOF current pulse for various values of energetic disorder in linear scale. Double logarithmic representation is shown in the inset. Current pulses are normalized to the magnitude of the current at the plateau region. Arrow represents the increasing direction of energetic disorder. Simulations were performed at E=6.4x10$^5$V/cm and T=300K.

Figure 2. Variation in the value of $W$ (Calculated from TOF current pulses simulated at E=6.4x10$^5$V/cm and T=300K.) for various value of energetic disorder. Inset shows the variation of mobility with energetic disorder.

Figure 3. Temporal evolution of $D_z$ of carriers for various value of energetic disorder. Simulations performed at E=6.4x10$^5$V/cm and T=300K.

Figure 4. Simulated TOF current pulses for various concentrations of ordered regions in linear scale. Double logarithmic representation is shown in the inset. Current pulses are normalized to the magnitude of the current at the plateau region. Arrow represents the increasing direction of concentrations of ordered regions. Simulations performed at E=6.4x10$^5$V/cm and T=300K.

Figure 5. Variation in the value of $W$ (Calculated from TOF current pulses simulated at E=6.4x10$^5$V/cm and T=300K.) for various concentrations of ordered regions. Inset shows the variation of mobility for various concentrations of ordered regions.

Figure 6. Temporal evolution of $D_z$ for various concentrations of ordered regions. Simulation performed at E=6.4x10$^5$V/cm and T=300K.

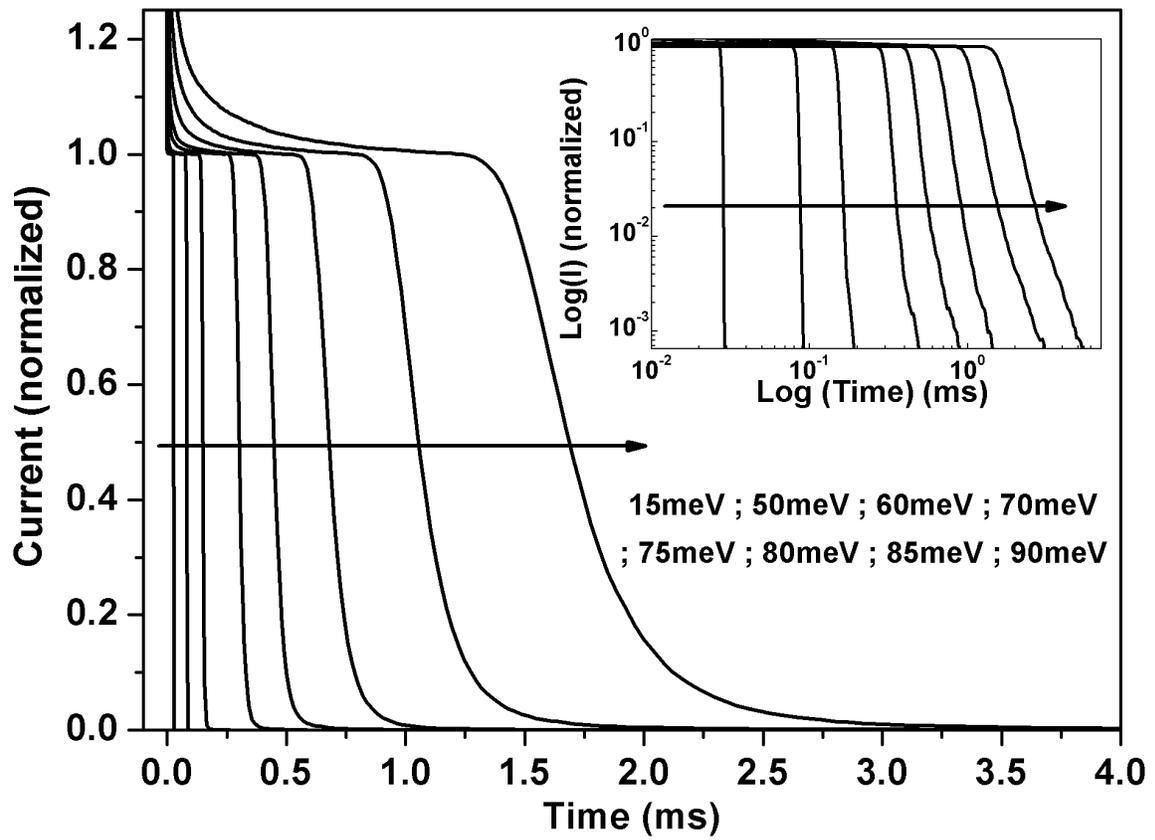

Figure 1

15meV ; 50meV ; 60meV ; 70meV ; 75meV ; 80meV ; 85meV ; 90meV

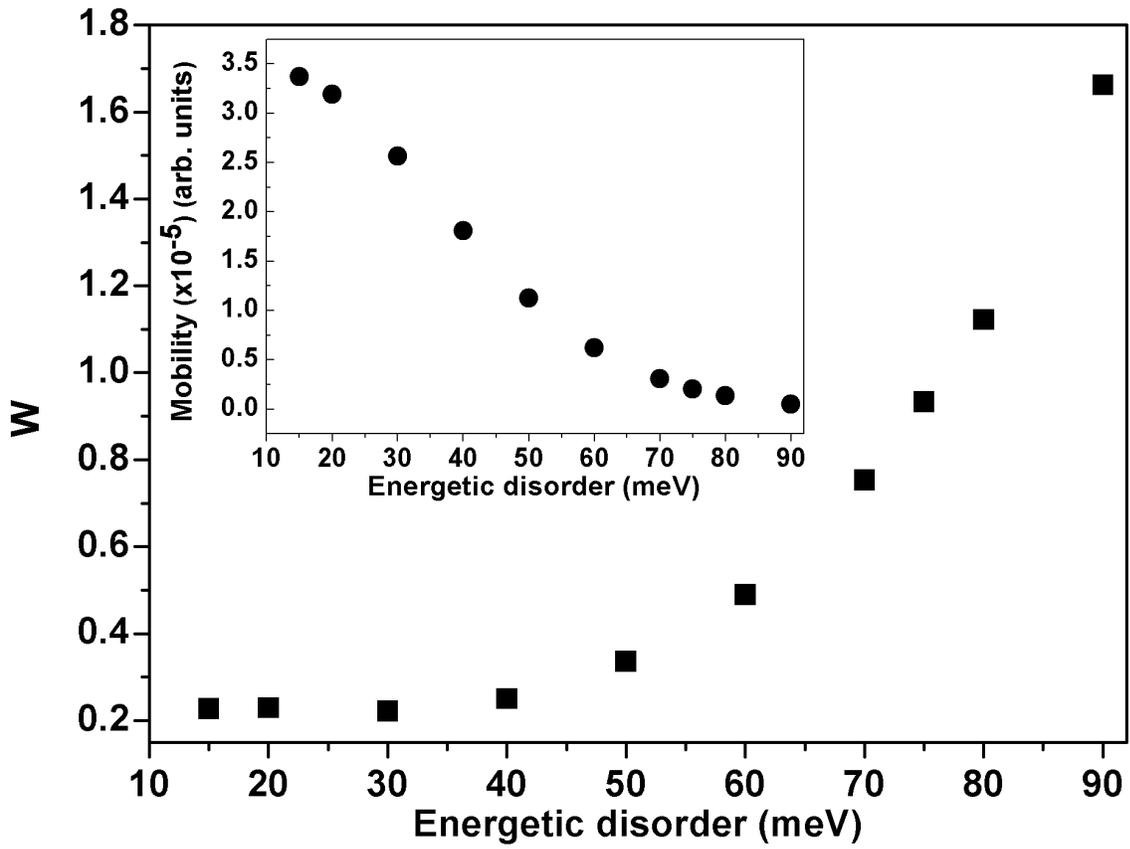

Figure 2

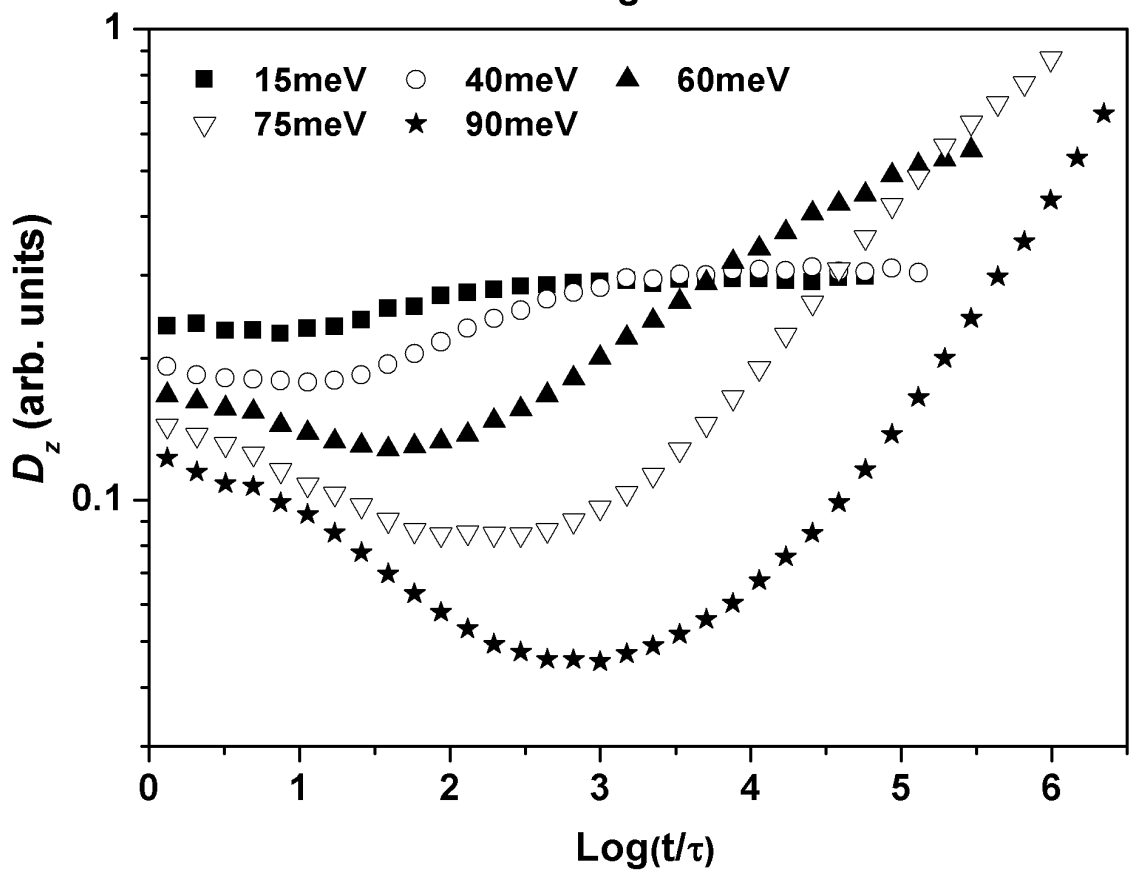

Figure 3

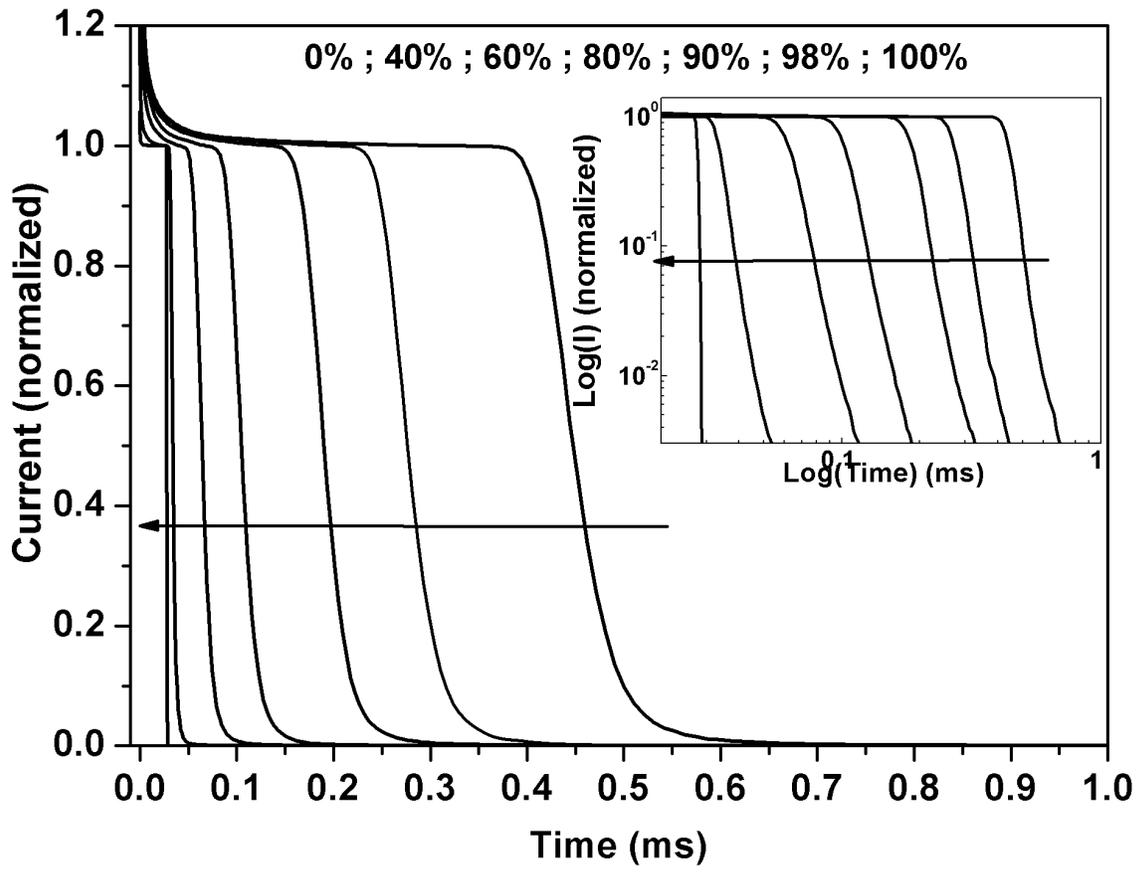

Figure 4

0% ; 40% ; 60% ; 80% ; 90% ; 98% ; 100%

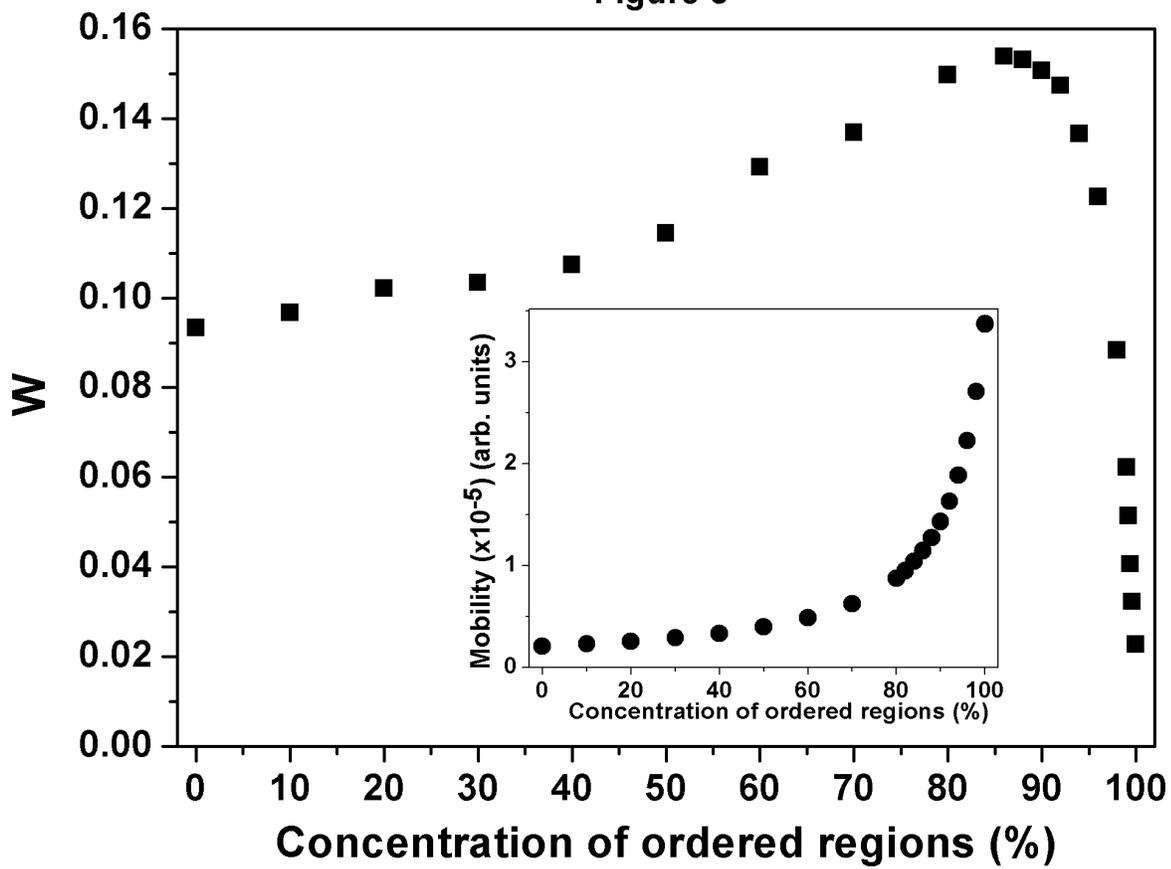

Figure 5

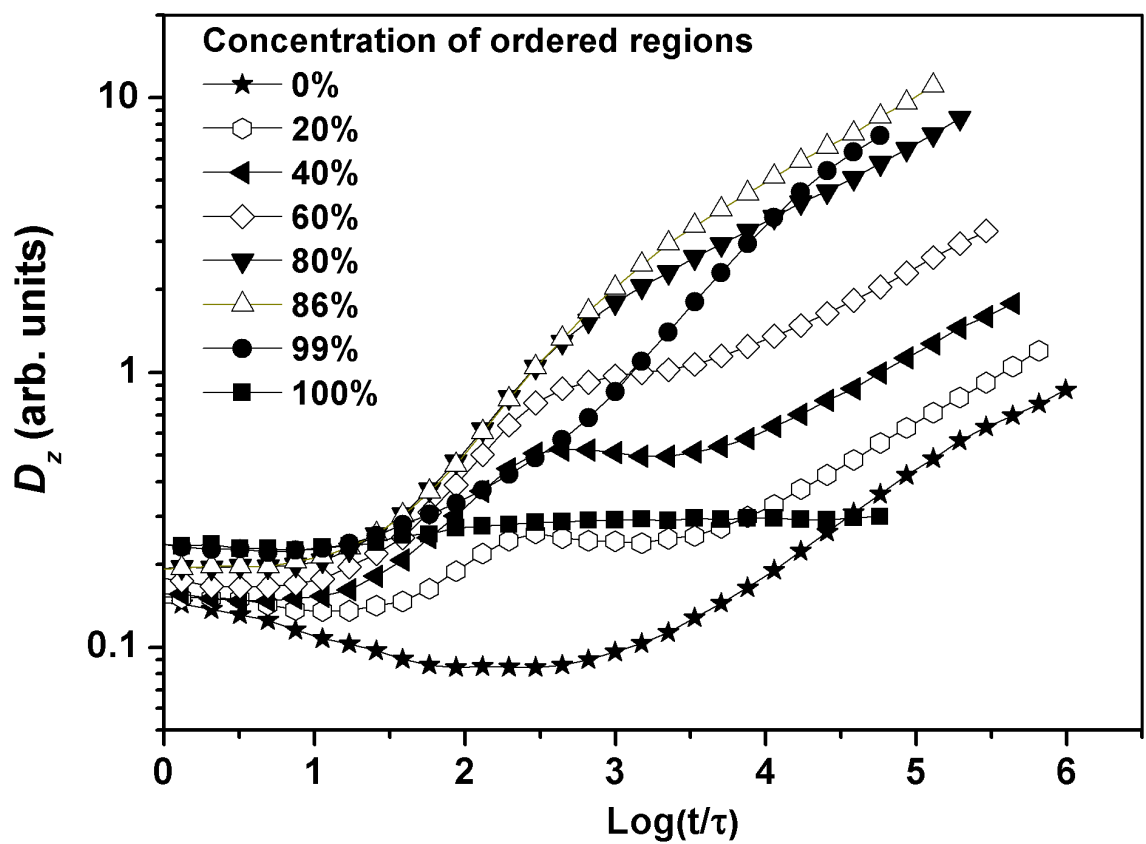